\documentclass{article} 
\usepackage{times}

\usepackage[hidelinks]{hyperref}
\usepackage{url}
\usepackage{booktabs}       
\usepackage{amsfonts}       
\usepackage{nicefrac}       
\usepackage{microtype}      
\usepackage{amsmath}
\usepackage{color}
\usepackage{multirow}
\usepackage{graphicx}
\usepackage{natbib}

\newcommand{\expect}{\mathbb{E}}
\renewcommand{\eqref}[1]{Eq.~(\ref{#1})}

\newcommand{\figref}[1]{Fig.~\ref{#1}}
\newcommand{\secref}[1]{Section~\ref{#1}}
\newcommand{\appref}[1]{Appendix~\ref{#1}}
\newcommand{\tabref}[1]{Table~\ref{#1}}

\newcommand{\blue}[1]{{\textcolor{blue}{#1}}}

\renewcommand{\cite}{\citep}
\newcommand{\reals}{\mathbb{R}}

\DeclareMathOperator*{\argmax}{argmax}
\newcommand{\dropout}{\mathit{dropout}}
\newcommand{\enc}{\mathit{enc}}
\newcommand{\dec}{\mathit{dec}}
\newcommand{\go}{\mathit{go}}
\newcommand{\hinge}{\mathit{hinge}}
\newcommand{\xent}{\mathit{xent}}

\makeatletter
\renewcommand{\paragraph}{%
  \@startsection{paragraph}{4}%
  {\z@}{1.0ex \@plus 1ex \@minus .2ex}{-1em}%
  {\normalfont\normalsize\bfseries}%
}
\makeatother

\newcommand*\samethanks[1][\value{footnote}]{\footnotemark[#1]}

\date{}

\title{Seq2Slate: Re-ranking and Slate Optimization with RNNs}

\author{
Irwan Bello\thanks{Corresponding authors: \href{mailto:ibello@google.com}{ibello@google.com}, \href{mailto:meshi@google.com}{meshi@google.com}}
\and
Sayali Kulkarni
\and
Sagar Jain
\and
Craig Boutilier
\and
Ed Chi
\and
Elad Eban
\and
Xiyang Luo
\and
Alan Mackey
\and
Ofer Meshi\samethanks \\
 \\
Google
}

\begin{document}

\maketitle

\begin{abstract}
\noindent
Ranking is a central task in machine learning and information retrieval.
In this task, it is especially important to present the user with a slate of items that is appealing as a whole.
This in turn requires taking into account interactions between items, since intuitively, placing an item on the slate affects the decision of which other items should be placed alongside it.
In this work, we propose a sequence-to-sequence model for ranking called \emph{seq2slate}.
At each step, the model predicts the next ``best'' item to place on the slate given the items already selected.
The sequential nature of the model allows complex dependencies between the items to be captured directly in a flexible and scalable way.
We show how to learn the model end-to-end from weak supervision in the form of easily obtained click-through data. We further demonstrate the usefulness of our approach in experiments on standard ranking benchmarks as well as in a real-world recommendation system.
\end{abstract}

\section{Introduction}
Ranking a set of candidate items is a central task in machine learning and information retrieval.
Many existing ranking systems are based on pointwise estimators, where the model assigns a score to each item in a candidate set and the resulting \emph{slate} is obtained by sorting the list according to item scores \cite{liu2009letor}.
Such models are usually trained from click-through data to optimize an appropriate loss function \cite{joachims2002}.
This simple approach is computationally attractive as it only requires a sort operation over the candidate set at test (or serving) time, and can therefore scale to large problems.
On the other hand, in terms of modeling, pointwise rankers cannot easily express dependencies between ranked items. In particular, the score of an item (e.g., its probability of being clicked) often depends on the other items in the slate and their joint placement.
Such interactions between items can be especially dominant in the common case where display area is limited or when strong position bias is present, so that only a few highly ranked items get the user's attention.
In this case it may be preferable, for example, to present a \emph{diverse} set of items at the top positions of the slate in order to cover a wider range of user interests.
Conversely, presenting multiple items with similar attributes may create ``synergies'' by drawing attention to
the collection, amplifying user response beyond that of any individual item. 

A significant amount of work on learning-to-rank does consider interactions between ranked items when \emph{training} the model.
In \emph{pairwise} approaches a classifier is trained to determine which item should be ranked first within a pair of items \cite[e.g.,][]{ordinal_svm,joachims2002,RankNet2005}.
Similarly, in \emph{listwise} approaches the loss depends on the full permutation of items \cite[e.g.,][]{cao2007listwise,yue2007map}.
Although these losses consider inter-item dependencies, the ranking function itself is pointwise, so at inference time the model still assigns a score to each item which does not depend on scores of other items (i.e., an item's score will not change if it is placed in a different set).

There has been some work on trying to capture interactions between items in the ranking scores themselves \cite[e.g.,][]{qin2008,qin2009,zhu2014,rosenfeld2014,dokania2014ranking,Borodin2017,GroupwiseRankingArXiv}.
Such approaches can, for example, encourage a pair of items to appear next to (or far from) each other in the resulting ranking.
Approaches of this type often assume that the relationship between items takes a simple form (e.g., submodular \cite{Borodin2017}) in order to obtain tractable inference and learning algorithms. Unfortunately, this comes at the expense of the model's expressive power.
Alternatively, greedy or approximate procedures can be used at inference time, though this often introduces approximation errors, and many of these procedures are still computationally expensive \cite[e.g.,][]{rosenfeld2014}.

More recently, neural architectures have been used to extract representations of the entire set of candidate items for ranking, thereby taking into consideration all candidates when assigning a score for each item \cite{Mottini2017kdd,ai2018sigir}.
This is done by an encoder which processes all candidate items sequentially and produces a compact representation,
followed by a scoring step in which pointwise scores are assigned based on this joint representation.
This approach can in principle model rich dependencies between ranked items, however its modeling requirements are quite strong. In particular, all the information about interactions between items needs to be stored in the intermediate compact representation and extracted in one-shot when scoring (decoding).

Instead, in this paper we propose a different approach by applying \emph{sequential decoding}, which assigns item scores conditioned on previously chosen items.
Our decoding procedure lets the score of an item change depending on the items already placed in previous positions.
This in turn allows the model to account for high-order interactions in a natural and scalable manner.
Moreover, our approach is purely data-driven so the model can adapt to various types of inter-item dependencies, including
synergies---where items appearing together contribute to their joint appeal,
and interference---where items decrease each other's appeal.
In particular, we apply a \emph{sequence-to-sequence (seq2seq)} model \cite{Sutskever2014} to the ranking task, where the input is the list of candidate items and the output is the resulting ordering.
Since the output sequence corresponds to ranked items on the slate, we call this approach \emph{sequence-to-slate}, or in short \emph{seq2slate}.

\begin{figure*}[t!]
    \centering
    \includegraphics[width=0.9\textwidth]{{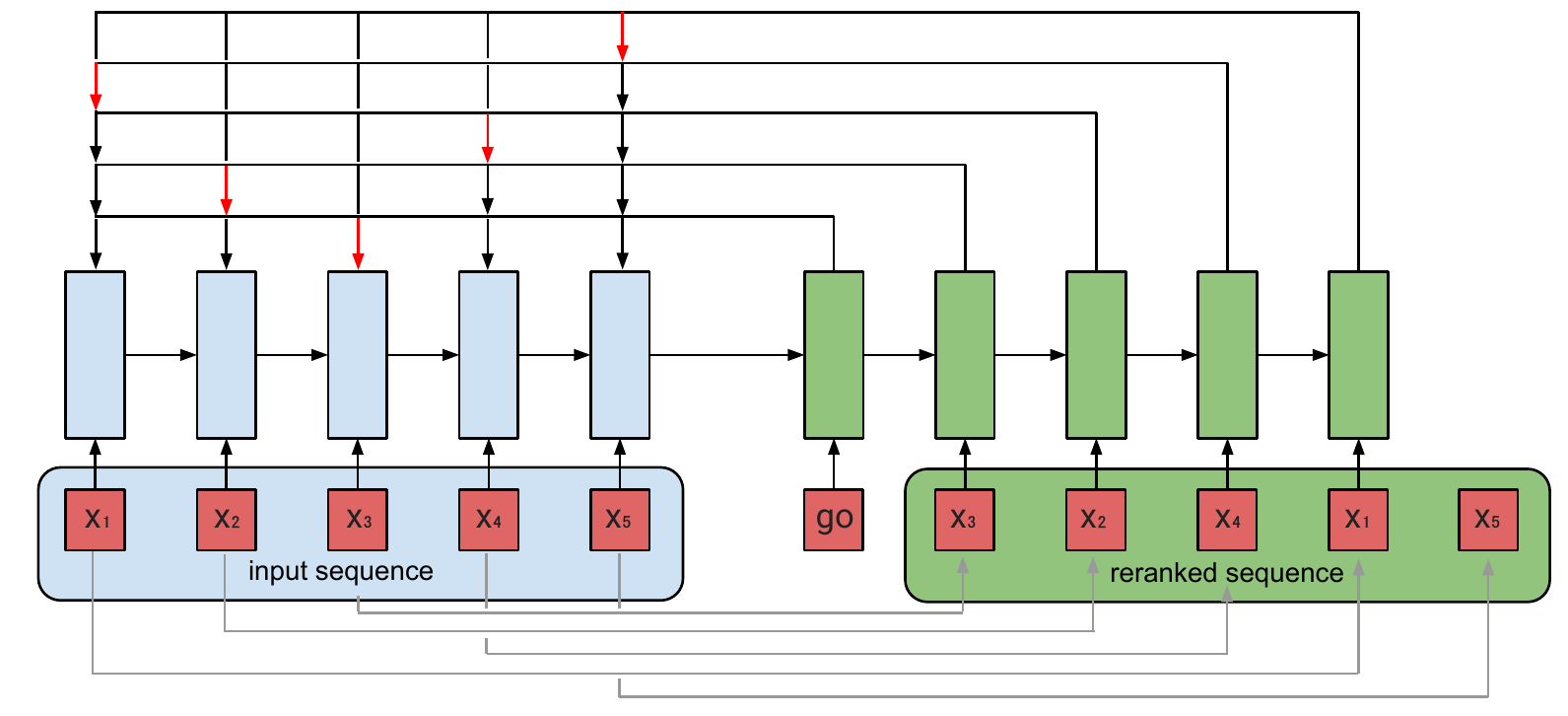}} 
    \caption{The seq2slate pointer network architecture for ranking.}
    \label{fig:ptrnet}
\end{figure*}

To address the seq2seq problem, we build on the recent success of \emph{recurrent neural networks (RNNs)} in a wide range of applications \cite[e.g.,][]{Sutskever2014}.
This allows us to use a deep model to capture rich dependencies between ranked items, while keeping the computational cost of inference manageable.
More specifically, we use \emph{pointer networks}, which are seq2seq models with an attention mechanism for pointing at positions in the input \cite{vinyals2015}.
We show how to train the network end-to-end to optimize several commonly used ranking measures.
To this end, we adapt RNN training to use weak supervision in the form of click-through data obtained from logs, instead of relying on ground-truth rankings, which are much more expensive to obtain.
Finally, we demonstrate the usefulness of the proposed approach in a number of learning-to-rank benchmarks and in a large-scale, real-world recommendation system.

\section{Ranking as Sequence Prediction}
\label{sec:model}

The \emph{ranking problem} is that of computing a ranking of a set of items (or ordered list or \emph{slate}) given some query or context.
We formalize the  problem as follows.
Assume a set of $n$ items, each represented by a feature vector $x_i\in\reals^m$ (which may depend on a query or context).%
\footnote{$x_i$ can represent either raw inputs or learned embeddings.}
Let $\pi\in\Pi$ denote a permutation of the items, where each $\pi_j \in \{1,\ldots,n\}$ denotes the index of the item in position $j$, for example, $\pi=(3,1,2,4)$ for $n=4$.
Our goal is to predict an ``optimal'' output ranking $\pi$ given the input items $x$.
For instance, given a specific user query, we might want to return an ordered set of music recommendations from a set of candidates that maximizes some measure of user engagement (e.g., number of tracks played).

In the seq2seq framework, the probability of an output permutation, or slate,
given the inputs is expressed as a product of conditional probabilities according to the chain rule:
\begin{equation}
\label{eq:chain_rule}
p(\pi|x) = \prod_{j=1}^n p(\pi_j|\pi_1,\ldots,\pi_{j-1}, x) ~,
\end{equation}
This expression is completely general and does not make any conditional independence assumptions.
In our case, the conditional $p(\pi_j|\pi_{<j},x)\in\Delta^n$ (a point in the $n$-dimensional simplex) models the probability of any item being placed at the $j$'th position in the ranking given the items already placed at previous positions.
For brevity, we have denoted the prefix permutation $\pi_{<j}=(\pi_1,\ldots,\pi_{j-1})$.
Therefore, this conditional can exactly capture \emph{all} high-order dependencies between items in the ranked list, including those due to diversity, similarity or other interactions.

Our setting is somewhat different than a standard seq2seq setting in that the output vocabulary 
is not fixed.
In particular, unlike in e.g., machine translation, the same index (position) is populated by different items in different instances (queries).
The vocabulary size $n$ itself may also vary per instance in the common case where the number of items to rank can change.
This is precisely the problem addressed by \emph{pointer networks}, which we review next.

\subsection*{Pointer-Network Architecture for Ranking}
We employ the \emph{pointer-network architecture} of \citet{vinyals2015} to model the conditional $p(\pi_j|\pi_{<j},x)$. A pointer network uses non-parametric softmax modules, akin to the attention mechanism of \citet{Bahdanau2015}, and learns to point to items in its input sequence rather than predicting an index from a fixed-sized vocabulary.

Our \emph{seq2slate} model, illustrated in \figref{fig:ptrnet},
consists of two \emph{recurrent neural networks} (RNNs): an encoder and a decoder, both of which use Long Short-Term Memory (LSTM) cells~\citep{lstm97}.
At each encoding step $i\leq n$, the encoder RNN reads the input vector $x_i$ and outputs a $\rho$-dimensional vector $e_i$, thus transforming the input sequence $\{x_i\}_{i=1}^n$ into a sequence of latent memory states $\{e_i\}_{i=1}^n$.
These latent states can be seen as a compact representation of the entire set of candidate items.
At each decoding step $j$, the decoder RNN outputs a $\rho$-dimensional vector $d_j$ which is used as a query in the attention function.
The attention function takes as input the query $d_j \in \mathbb{R}^{\rho}$ and the set of latent memory states computed by the encoder $\{e_i\}_{i=1}^n$ and produces a probability distribution over the next item to include in the output sequence as follows:
{\small 
\begin{align}
\label{eq:attn}
    &s^j_i = v^{\top} \tanh{\left( W_{enc} \cdot e_i + W_{dec} \cdot d_j \right)} \\
    &p_{\theta}(\pi_j = i |\pi_{<j},x) ~\equiv p_i^j =
    \begin{cases}
        e^{s_i^j} / \sum_{k\notin\pi_{<j}} e^{s_k^j} & \text{if } i\notin \pi_{<j} \\
        0 & \text{if } i\in\pi_{<j}
    \end{cases}
                                      \qquad.  \nonumber
\end{align}
}%
Here $W_{\enc}, W_{\dec} \in \mathbb{R}^{\rho \times \rho}$ and $v \in \mathbb{R}^{\rho}$ are learned parameters in our network, denoted collectively by parameter vector $\theta$,
and $s^j_i$ are \emph{scores}
associated with placing item $i$ in
position $j$. The probability $p_i^j = p_{\theta}(\pi_j = i |\pi_{<j},x)$,
is obtained via a softmax over the remaining items and represents the degree to which the model points to input $i$ at decoding step $j$.
In order to output a permutation, the probabilities $p_i^j$ are set to 0 for items $i$ that already appear on the slate. %
Once the next item $\pi_j$ is selected, typically greedily or by sampling (see below), its embedding $x_{\pi_j}$ is fed as input to the next decoder step.
This way the decoder states hold information on the items already placed on the slate.
The input to the first decoder step is a learned $m$-dimensional vector, denoted as `$\go$' in \figref{fig:ptrnet}.

We note the following.
\textbf{(i)} Our formulation using sequential decoding lets the score of items (i.e., $p_i^j$) change depending on items
previously placed on the slate, thereby allowing the model to account for high-order interactions in a natural way.
\textbf{(ii)} The model makes no explicit assumptions about the type of interactions between items. If the learned conditional in \eqref{eq:attn} is close to the true conditional in \eqref{eq:chain_rule}, then the model can capture rich interactions---including diversity, similarity or others.
Hence, our approach is data-driven rather than modeling specific types of interactions
(such as multinomial logit), which is a key advantage.
We demonstrate the benefits of this flexibility in our experiments (\secref{sec:experiments}).
\textbf{(iii)} The probability $p_\theta(\pi | x)$ is differentiable (in $\theta$) for any fixed permutation $\pi$, which allows gradient-based learning (see \secref{sec:training}).
\textbf{(iv)} The computational cost of inference, dominated by the sequential decoding procedure, is $O(n^2)$, which is standard in seq2seq models with attention.
We also consider a computationally cheaper single-step decoder with linear cost $O(n)$, which outputs a single vector $p^1=p_\theta(\pi_1=\cdot|x)$ (see \eqref{eq:attn}), from which we obtain $\pi$ by sorting the values---similar to the approach taken in \cite{Mottini2017kdd,ai2018sigir}); we compare both approaches below.

Previous studies have shown that the order in which the input is processed can significantly affect the performance of sequential models
\cite{vinyals2016,nam2017multilabel,ai2018sigir}.
For this reason, we will assume here the availability of a base (or ``production'') ranker with which the input sequence is ordered (e.g., a simple pointwise method that ignores the interactions we seek to model), and view the output of our model as a \emph{re-ranking} of the items.
In many real systems such base ranker is readily available. For example, the candidate set may be chosen from a huge item repository by an upstream model. Often candidate generator scores are available and can be used to obtain a base ranking via a simple sort. In this case we obtain the base ranking almost for free, as byproduct of candidate generation.
Importantly, using a base ranker and focusing on re-ranking allows our seq2slate model to direct its modeling capacity at interactions between items rather than individual items.

\section{\label{sec:training}Training with Click-Through Data}

We now turn to the task of training the seq2slate model from data.
A typical approach to learning in ranking systems is to run an existing ranker ``in the wild'' and log click-through data, which are then used to train an improved ranking model. This type of training data is relatively inexpensive to obtain, in contrast to
human-curated labels such as relevance scores, ratings, or full rankings \cite{joachims2002}.

Formally, each training example consists of a sequence of items $x=\{x_1,\ldots,x_n\}$, with $x_i\in\reals^m$
and binary labels $y=(y_1,\ldots,y_n)$, 
with $y_i\in\{0,1\}$, representing user feedback (e.g., click/no-click).
Our approach can be easily extended to more informative feedback, such as the level of user engagement with the chosen item (e.g., time spent), but to simplify the presentation we focus on the binary case.
Our goal is to learn the parameters $\theta$ of $p_{\theta}(\pi_j |\pi_{<j},x)$ (\eqref{eq:attn}) such that permutations $\pi$ corresponding to ``good'' rankings are assigned high probabilities.
Various performance measures ${\mathcal{R}}(\pi,y)$ can be used to evaluate the quality of a permutation $\pi$ given the labels $y$, for example, mean average precision (MAP), precision at $k$, or normalized discounted cumulative gain at $k$ (NDCG@k). Generally speaking, permutations where the positive labels rank higher are considered better.

In the standard seq2seq setting, models are trained to maximize the likelihood of a target sequence of tokens given the input, which can be done by maximizing the likelihood of each target token given the previous target tokens using \eqref{eq:chain_rule}.
In this case, the model is typically fed the ground-truth tokens as inputs to the next prediction step during training, an approach known as \emph{teacher forcing} \cite{Williams89TeacherForcing}.
Unfortunately, this approach cannot be applied in our setting since we only have access to weak supervision in the form of labels $y$ (e.g., clicks), rather than ground-truth permutations.
Instead, we next show how the seq2slate model can be trained directly from the labels $y$.

\subsection{Training Using \textsc{Reinforce}}
\label{sec:policy_gradient}
One viable approach, which has been applied successfully in related tasks~\citep{bello2017nco,zhongSQL}, is to use \emph{reinforcement learning (RL)} to directly optimize for the ranking measure ${\mathcal{R}}(\pi,y)$. In this setup, the objective is to maximize the expected ranking metric obtained by sequences sampled from our model:
\[
\max_\theta ~~ \mathbb{E}_{\pi \sim p_{\theta}(.|x)} [{\mathcal{R}}(\pi,y)]~.
\]
One can use policy gradients and stochastic gradient ascent to optimize $\theta$. 
The gradient is formulated using the popular \textsc{reinforce} update~\citep{reinforce}:
{\small
\begin{align}
\label{eq:policy_grad}
  \nabla_\theta \mathbb{E}_{\pi \sim p_{\theta}(.|x)} [{\mathcal{R}}(\pi,y)] &= 
  \mathbb{E}_{\pi \sim p_{\theta}(.\mid x)}\Big[{\mathcal{R}}(\pi,y) \nabla_\theta
    \log{p_{\theta}(\pi \mid x)}\Big]~.
\end{align}
}%
This can be approximated via Monte-Carlo sampling as follows:
{\small
\begin{align}
\label{eq:policy_grad_samples}
&\approx \frac{1}{B} \sum_{k=1}^B \Big({\mathcal{R}}(\pi[k],y[k])-b_\mathcal{R}(x[k])\Big) \nabla_\theta \log{p_{\theta}(\pi[k] \mid x[k])}~,
\end{align}
}%
where $k$ indexes ranking instances in a batch of size $B$, the $\pi[k]$ are permutations drawn from the model $p_\theta$, and $b_\mathcal{R}(x)$ denotes a baseline function that estimates the expected rewards in order to reduce variance.

\subsection{Supervised Training}
Policy gradient methods like \textsc{reinforce} are known to induce challenging optimization problems and can suffer from sample inefficiency and difficult credit assignment. As an alternative, we propose \emph{supervised learning} using the labels $y$.
In particular, rather than waiting until the end of the output sequence as in RL above, we can give feedback to the model at each decoder step.

Consider the first step, and recall that
the model assigns a score $s_i$ to each item in the input (see~\eqref{eq:attn}); to simplify notation we omit the 
position superscript $j$ for now.
Letting $s=(s_1,\ldots,s_n)$,
we define a per-step loss $\ell(s,y)$ which essentially acts as a multi-label classification loss with labels $y$ as ground truth.
Two natural, simple choices for $\ell$ are cross-entropy loss and hinge loss:
\begin{align}
\label{eq:losses}
\ell_{\xent}(s,y) &= -\sum_i \hat{y}_i \log p_i \\
\ell_{\hinge}(s,y) &= \max\{0, 1 - \min_{i:y_i=1}s_i + \max_{j:y_j=0}s_j\} ~, \nonumber
\end{align}
where $\hat{y}_i=y_i/\sum_j y_j$, and $p_i$ is a softmax of $s$, as in \eqref{eq:attn}.
Intuitively, with cross-entropy loss we try to assign high probabilities to positive labels \cite[see also][]{kurata2016}, 
while hinge loss is minimized when scores of items with positive labels are higher than scores of those with negative labels.
Notice that both losses are convex functions of the scores $s$.
To improve convergence, we consider a smooth version of the hinge loss where the maximum and minimum are replaced by their smooth counterparts:
$\texttt{smooth-max}(s;\gamma)=\frac{1}{\gamma} \log \sum_i e^{\gamma s_i}$ (and smooth minimum is defined similarly, using $\min_i(s_i)=-\max_i(-s_i)$).
Finally, we point out that any standard surrogate loss for ranking can be used as the per-step loss $\ell(s,y)$, including losses that depend on non-binary labels $y$, such as relevance scores.

As mentioned above, a main difference of seq2slate from previous approaches is its use of sequential decoding. This does complicate the training of the model  somewhat relative to the
the case of one-shot decoding \cite{Mottini2017kdd,ai2018sigir}.
Specifically, if we simply apply a per-step loss from \eqref{eq:losses} to all steps of the output sequence while reusing the labels $y$ at each step, then the loss is invariant to 
the resulting output permutation (i.e., predicting a positive item at the beginning of the sequence has the same cost as predicting it at the end).
Instead, in order to train a seq2slate model we let the loss $\ell$ at each decoding step $j$ ignore the items already chosen, so no further loss is incurred after a label is predicted correctly.
In particular, for a \emph{fixed} permutation $\pi$, define the \emph{sequence loss}:
\begin{align}
{\mathcal{L}}_\pi(S,y) = \sum_{j=1}^n w_j~ \ell_{\pi_{<j}}(s^j,y) ~,
\label{eq:sequence_loss}
\end{align}
where $S=\{s^j\}_{j=1}^n$ are the model scores (see~\eqref{eq:attn}), and each $s^j = (s^j_1,\ldots,s^j_n)$ is the item-score vector for position $j$.
In the sequel we will also use the abbreviation: ${\mathcal{L}}_\pi(\theta)\equiv{\mathcal{L}}_\pi(S(\theta), y)$.
Importantly, the per-step loss $\ell_{\pi_{<j}}(s^j,y)$ depends only on the indices in $s^j$ and $y$ which are not in the prefix $\pi_{<j}$ (cf.~\eqref{eq:losses}).
Including a per-step weight $w_j$ can encourage better performance earlier in the sequence. For example, we might set $w_j=1/\log(j+1)$ (along the lines of DCG).
Alternatively, if optimizing for a particular slate size $k$ is desired, one can use the weights to restrict this loss to just the first $k$ output steps.

We note that the loss above differs from the actual ranking measures used in evaluation (i.e., MAP, NDCG@k, etc.).
On the other hand, any permutation that places the positive labels at the first positions gets 0 loss and optimizes all ranking measures, so in that sense the losses are aligned. This situation is quite common for surrogate losses in machine learning.

Using the definition of the sequence loss above, our goal is to optimize the expected loss:
\[
\min_\theta ~ \expect_{\pi\sim p_\theta(\cdot|x)}[{\mathcal{L}}_\pi(\theta)]~,
\]
where
\begin{equation}
\label{eq:expected_loss}
\expect_{\pi\sim p_\theta(\cdot|x)}[{\mathcal{L}}_\pi(\theta)] = \sum_\pi p_\theta(\pi|x) {\mathcal{L}}_\pi(\theta)~.
\end{equation}
This corresponds to sampling the permutation $\pi$ according to the model, where $\pi_j$ is drawn from $p_\theta(\cdot|\pi_{<j},x)$ for each position $j$.
For completeness, we derive the expected loss as a function of the model scores $S$ in \appref{app:expected_loss}.

Notice that the expected loss in \eqref{eq:expected_loss} is differentiable everywhere since both $p_\theta(\pi|x)$ and ${\mathcal{L_\pi}}(\theta)$ are differentiable for any permutation $\pi$.
In this case, the gradient is formulated as:
{\small
\begin{align}
\nabla_\theta \expect_\pi[{\mathcal{L}}_\pi(\theta)] =&~ \nabla_\theta \sum_\pi p_\theta(\pi|x) {\mathcal{L}}_\pi(\theta) \nonumber\\
=&~
\sum_\pi \left[ (\nabla_\theta p_\theta(\pi|x)) {\mathcal{L}}_\pi(\theta) + p_\theta(\pi|x) (\nabla_\theta {\mathcal{L}}_\pi(\theta)) \right] \nonumber\\
=&~
\expect_{\pi\sim p_\theta(\cdot|x)} \left[ {\mathcal{L}}_\pi(\theta) \cdot \nabla_\theta \log p_\theta(\pi|x) +
\nabla_\theta {\mathcal{L}}_\pi(\theta) \right]~,
\label{eq:sampling_gradient}
\end{align}
}%
which can be approximated from samples by:
{\small
\begin{align}
\approx~ \frac{1}{B} \sum_{k=1}^B \Bigg[ & \Big({\mathcal{L}}_{\pi[k]}(S(\theta),y[k]) - b_\mathcal{L}(x[k])\Big) \nabla_\theta \log p_{\theta}(\pi[k] \mid x[k]) \nonumber\\
&+ \nabla_\theta {\mathcal{L}}_{\pi[k]}(S(\theta),y[k]) \Bigg] ~.
\label{eq:sampling_gradient_samples}
\end{align}
}%
Here $b_\mathcal{L}(x[k])$ is a baseline that approximates ${\mathcal{L}}_{\pi[k]}(\theta)$, introduced for variance reduction.
This gradient is analogous to the \textsc{reinforce} update from \eqref{eq:policy_grad}--(\ref{eq:policy_grad_samples}), but where the loss ${\mathcal{L}}$ subsumes the role of the reward ${\mathcal{R}}$.
Notice, however, that since the loss depends on the model parameters $\theta$ while the reward does not, the resulting update is quite different.
Specifically, applying stochastic gradient descent intuitively decreases
the probability of drawing samples with high losses (left term in \eqref{eq:sampling_gradient}), as in \textsc{reinforce}, but in addition also reduces the loss of any sample (right term in \eqref{eq:sampling_gradient}), which differs from \textsc{reinforce} \cite[see also][Eq.~(4)]{schulman2015}.

\subsubsection*{\bf Greedy Decoding}
In many seq2seq applications, using greedy decoding at test time performs better than sampling from the model \cite[e.g.,][]{Ranzato2016,searnn2018leblond}.
Therefore, it makes sense to also consider training the model using a greedy decoding policy, which is an alternative approach to sampling (cf.~\eqref{eq:sampling_gradient_samples}).
The greedy policy consists of selecting the item that maximizes $p_\theta(\cdot|\pi_{<j},x)$ at every step $j$.
The resulting permutation $\pi^*$ then satisfies $\pi_j^* = \argmax_i ~ p_\theta(\pi_j=i|\pi^*_{<j},x)$ and our loss simply becomes ${\mathcal{L}}_{\pi^*}(\theta)$.
Unlike the sampling-based loss in \eqref{eq:expected_loss}, the greedy policy loss is not continuous everywhere since a small change in the scores $S$ may result in a jump between permutations $\pi^*$, and therefore a jump in the value of ${\mathcal{L}}_{\pi^*}(\theta)$.
Specifically, the loss is non-differentiable when any $s^j$ has multiple maximizing arguments. Outside this measure-zero subspace, the loss is continuous (almost everywhere), and the gradient is well-defined.

For both training policies (sampling and greedy), we minimize the loss via stochastic gradient descent over mini-batches in an \emph{end-to-end} fashion.

\section{Experimental Results}
\label{sec:experiments}
We evaluate the performance of our seq2slate model on a collection of ranking tasks.
In \secref{sec:letor_experiments} we use learning-to-rank benchmark
data to study the behavior of the model.
We then apply our approach to a large-scale commercial recommendation system and report the results in \secref{sec:real_experiments}.

\paragraph{Implementation details}
We set hyperparameters of our model to values inspired by the literature.
All experiments use mini-batches of $128$ training examples and LSTM cells with $128$ hidden units. We train our models with the Adam optimizer~\citep{adam} and an initial learning rate of $0.0003$ decayed every $1000$ steps by a factor of $0.96$.  Network parameters are initialized uniformly at random in $[-0.1, 0.1]$. To improve generalization, we regularize the model by using dropout with probability of dropping $p_{\dropout}=0.1$ and L2 regularization with a penalty coefficient $\lambda=0.0003$.
Unless specified otherwise, all results use supervised training with cross-entropy loss $\ell_{\xent}$ and the sampling policy. At inference time, we report metrics for the greedy policy.
We use an exponential moving average with a decay rate of $0.99$ as the baseline functions $b_\mathcal{R}(x)$ and $b_\mathcal{L}(x)$ in \eqref{eq:policy_grad_samples} and (\ref{eq:sampling_gradient_samples}), respectively. When training the seq2slate model with \textsc{reinforce}, we use ${\mathcal{R}}=\texttt{NDCG}@10$ as the reward function and do not regularize the model (since we observed no overfitting during training with the noisy policy gradients).
We also considered a bidirectional encoder RNN~\citep{bidir}, a stacked LSTM, and models with more hidden units, but found that these did not lead to significant improvements in our experiments.

\subsection{Learning-to-Rank Benchmarks}
\label{sec:letor_experiments}
To understand the behavior of the proposed model, we conduct experiments using two learning-to-rank datasets.
We use two of the largest publicly available benchmarks: the
\href{https://webscope.sandbox.yahoo.com/catalog.php?datatype=c}{\blue{Yahoo Learning to Rank Challenge}} data
 (set 1),%
\footnote{{\scriptsize \url{https://webscope.sandbox.yahoo.com/catalog.php?datatype=c}}}
and the \href{https://www.microsoft.com/en-us/research/project/mslr/}{\blue{Microsoft Web30k}} dataset.%
\footnote{{\scriptsize \url{https://www.microsoft.com/en-us/research/project/mslr/}}}
These datasets only provide feature vectors for each query-document pair, so all context (query) features are embedded within the item feature vectors themselves.

We adapt the procedure proposed by \citet{joachims2017} to generate click data. The original procedure is as follows:
first, a base ranker is trained from the raw data. We select this base ranker by training all models in the \href{https://sourceforge.net/p/lemur/wiki/RankLib/}{\blue{RankLib}} package,\footnote{{\scriptsize \url{https://sourceforge.net/p/lemur/wiki/RankLib/}}}
and choosing the one with the best performance on each data set (MART for Yahoo and LambdaMART for Web30k).
We generate an item ranking using the base model, which is then used to generate training data by simulating a user ``cascade'' model: a user observes each item
with decaying probability $1/i^\eta$, where $i$ is the base rank of the item and $\eta$ is a parameter of the generative model. This simulates
a noisy sequential scan by the user.
An observed item is clicked if its ground-truth relevance score is above a threshold (relevant: $\{2,3,4\}$, irrelevant: $\{0,1\}$), otherwise no click is generated.

Unfortunately, the original datasets only include a per-item relevance score, which is independent of the other items. This means that there are no direct high-order interactions between the clicks, and therefore the joint probability in \eqref{eq:chain_rule} is just $p(\pi|x) = \prod_{j=1}^n p(\pi_j|x)$. In this case a pointwise ranker is optimal so there would be no need for seq2slate.
Therefore, in order to introduce high-order dependencies, we augment the above procedure as follows, creating a generative process dubbed \emph{diverse-clicks}.
When observing a relevant item, the user will only click if it is not too similar to previously clicked items (i.e, diverse enough), thus reducing the total number of clicks. Similarity is defined as being in the
smallest $q$ percentile (i.e., $q=0.5$ is the median) of Euclidean distances between pairs of feature vectors within the same ranking instance: $D_{ij}=\|x_i-x_j\|$. 
We use $\eta=0$ (no decay, since clicks are sparse anyway due to the diversity term) and $q=0.5$.
We also discuss variations of this model below.
Since our focus is on modeling high-order interactions, all results reported in this section are w.r.t.~the generated binary labels and not the original relevance scores.

\begin{table*}[t!]
\begin{center}
{\small
\begin{tabular}{|c|ccc|ccc|}
\hline
\multirow{2}{*}{Ranker} & \multicolumn{3}{|c|}{Yahoo} & \multicolumn{3}{|c|}{Web30k} \\
{} & MAP & NDCG@5 & NDCG10 & MAP & NDCG@5 & NDCG@10 \\
\hline
seq2slate      & \bf{0.67} & \bf{0.69} & \bf{0.75} & \bf{0.51} & \bf{0.53} & \bf{0.59} \\
AdaRank             & 0.58 & 0.61 & 0.69 & 0.37 & 0.38 & 0.46 \\
Coordinate Ascent   & 0.49 & 0.51 & 0.59 & 0.31 & 0.33 & 0.39 \\
LambdaMART          & 0.58 & 0.61 & 0.69 & 0.42 & 0.46 & 0.52 \\
ListNet             & 0.49 & 0.51 & 0.59 & 0.43 & 0.47 & 0.53 \\
MART                & 0.58 & 0.60 & 0.68 & 0.39 & 0.42 & 0.48 \\
Random Forests      & 0.54 & 0.57 & 0.65 & 0.36 & 0.39 & 0.45 \\
RankBoost           & 0.50 & 0.52 & 0.60 & 0.24 & 0.25 & 0.30 \\
RankNet             & 0.54 & 0.57 & 0.64 & 0.43 & 0.47 & 0.53 \\
\hline
\end{tabular}
} 
\end{center}
\caption{Performance of seq2slate and other baselines on data generated with diverse-clicks.}
\label{tab:letor_diverse}
\end{table*}

Using the generated training data, we train both our seq2slate model and baseline rankers from the \href{https://sourceforge.net/p/lemur/wiki/RankLib/}{\blue{RankLib}} package:
AdaRank \cite{adarank},
Coordinate Ascent \cite{rank_coordinate_ascent},
LambdaMART \cite{lambdamart},
ListNet \cite{cao2007listwise},
MART \cite{mart},
Random Forests \cite{randomforests},
RankBoost \cite{rankboost},
RankNet \cite{RankNet2005}.
Some of these baselines use deep neural networks
(e.g., RankNet, ListNet),
so they are strong state-of-the-art models with comparable complexity to seq2slate.
The results in \tabref{tab:letor_diverse} show that seq2slate significantly outperforms all the baselines, suggesting that it can better capture and exploit the dependencies between items in the data.

\begin{figure}[t]
  \centering
  \includegraphics[width=150pt]{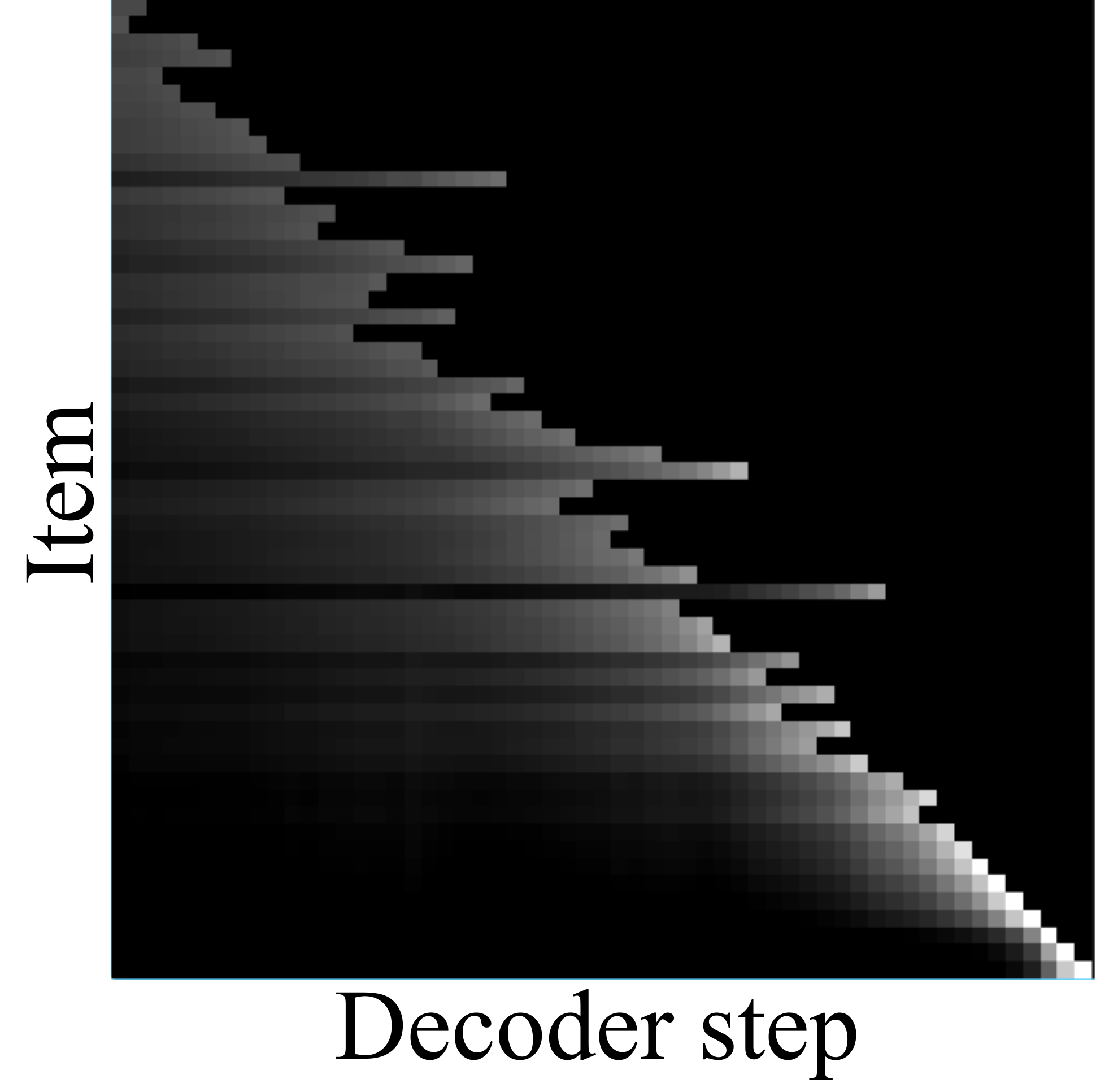}
  \caption{Visualization of attention probabilities on benchmark data. Intensities correspond to $p^j_i$ for each item $i$ in step $j$.}
  \label{fig:attention}
\end{figure}

To better understand the behavior of the model, we visualize the probabilities of the attention from \eqref{eq:attn} for one of the test instances in \figref{fig:attention}.
Interestingly, the model 
produces slates that are close to the input ranking, but with some items demoted to lower positions, presumably due to the interactions with previous items.

\begin{table*}[t!]
\begin{center}
{\footnotesize
\resizebox{0.95\textwidth}{!}{%
\begin{tabular}{|c|cccc|cccc|}
\hline
\multirow{2}{*}{Ranker} & \multicolumn{4}{|c|}{Yahoo} & \multicolumn{4}{|c|}{Web30k} \\
{} & MAP & NDCG@5 & NDCG@10 & rank-gain & MAP & NDCG@5 & NDCG@10 & rank-gain \\
\hline
seq2slate & \bf{0.67} & \bf{0.69} & \bf{0.75} & \bf{7.4} & \bf{0.51} & \bf{0.53} & \bf{0.59} & \bf{18.3} \\
Greedy policy      & 0.66 & 0.69 & 0.75 & 7.2 & 0.50 & 0.52 & 0.59 & 18.3 \\
smooth-hinge        & 0.66 & 0.69 & 0.75 & 7.1 & 0.49 & 0.51 & 0.58 & 17.9 \\
\textsc{reinforce}  & 0.66 & 0.68 & 0.75 & 5.7 & 0.44 & 0.47 & 0.53 & -0.5 \\
one-step decoder    & 0.66 & 0.69 & 0.75 & 6.4 & 0.49 & 0.51 & 0.58 & 16.5 \\
shuffled data       & 0.57 & 0.60 & 0.67 & -- & 0.40 & 0.40 & 0.48 & -- \\
base ranker (no-op) & 0.58 & 0.61 & 0.69 & 0 & 0.45 & 0.48 & 0.54 & 0 \\
\hline
\end{tabular}
} 
} 
\end{center}
\caption{Comparison of model and data variants for seq2slate on data generated with diverse-clicks.}
\label{tab:letor_variants}
\end{table*}

We next consider several variations of the generative model and of the seq2slate model itself.
Results are reported in \tabref{tab:letor_variants}.
The rank-gain metric per example is computed by summing the positions change of all positive labels in the re-ranking, and this is averaged over all examples (queries).

\paragraph{Comparison of training variants}
In \tabref{tab:letor_variants}, we compare the different training variants outlined in \secref{sec:training}, namely, cross entropy with the greedy or sampling policy, a smooth hinge loss with $\gamma=1.0$, and \textsc{reinforce}. We find that supervised learning with cross entropy generally performs best, with the smooth hinge loss doing slightly worse. Our weakly supervised training methods have positive rank gain on all datasets, meaning they improve over the base ranker. The results from \tabref{tab:letor_variants}
suggest that training with \textsc{reinforce} yields comparable results on Yahoo but significantly worse results on the more challenging Web30k dataset.
In terms of training time, \textsc{reinforce} needed 4X more time till convergence.
We find no significant difference in performance between relying on the greedy and sampling policies during training.

\paragraph{One-step decoding}
We compare seq2slate to the model which uses a single decoding step, referred to as \emph{one-step decoder} (see \secref{sec:model}).
In \tabref{tab:letor_variants} we see that this model has comparable performance to the sequential decoder.
One possible explanation for the comparable performance of the one-step decoder is that the interactions in our generated data are rather simple and can be effectively learned by the encoder. By contrast, in \secref{sec:real_experiments} we show that on more complex real-world data, sequential decoding can perform significantly better than one-step decoding.
In terms of runtime, we observed a 4X decrease in training time and a 3X decrease in inference time for the one-step decoder compared to sequential decoding (for the real-world data in \secref{sec:real_experiments} below, one-step decoding was 2.5X faster per iteration in both training and inference).
This suggests that when inference time is crucial, as in many real-world systems, one might prefer the faster single-shot option.
Having said that, we point out that even with sequential decoding the runtime was not a bottleneck in our case and we were able to train a seq2slate model on millions of examples in a couple of hours, and serve live traffic in $O(10)$ milliseconds.
For this reason we also did not make an effort to optimize the code, so the numbers above can probably be reduced significantly.

\paragraph{Sensitivity to input order}
Previous work suggests that the performance of seq2seq models is often sensitive to the order in which the input is processed \cite{vinyals2016,nam2017multilabel,ai2018sigir}. To test the sensitivity of seq2slate to the order in which items are processed, we consider the use of seq2slate without relying on the base ranker to order the input. Instead, items are fed to the model in random order.
Since learning the correct ranking from a single example may be hard, we generate multiple copies of each training example, each with a different randomly shuffled input order. Specifically, in \tabref{tab:letor_variants} we show results for 10 generated examples per original example under `shuffled data'.
The results show that the performance is indeed significantly worse in this case, which is consistent with previous studies. It suggests that reranking is an easier task than ranking from scratch.

\begin{table*}[t]
\begin{center}
\small{
\begin{tabular}{|c|ccc|ccc|}
\hline
\multirow{2}{*}{Ranker} & \multicolumn{3}{|c|}{Yahoo} & \multicolumn{3}{|c|}{Web30k} \\
{} & MAP & NDCG@5 & NDCG@10 & MAP & NDCG@5 & NDCG@10 \\
\hline
seq2slate           & 0.82 & 0.82 & 0.84 & \bf{0.44} & \bf{0.54} & \bf{0.50} \\
AdaRank             & 0.83 & 0.81 & 0.84 & 0.41 & 0.52 & 0.48 \\
Coordinate Ascent   & 0.83 & 0.82 & 0.85 & 0.39 & 0.47 & 0.44 \\
LambdaMART     & \bf{0.84} & \bf{0.83} & \bf{0.85} & 0.41 & 0.52 & 0.48 \\
ListNet             & 0.83 & 0.83 & 0.85 & 0.41 & 0.53 & 0.49 \\
MART                & 0.83 & 0.82 & 0.85 & 0.41 & 0.52 & 0.48 \\
Random Forests      & 0.83 & 0.82 & 0.84 & 0.40 & 0.48 & 0.45 \\
RankBoost           & 0.83 & 0.83 & 0.85 & 0.38 & 0.43 & 0.41 \\
RankNet             & 0.83 & 0.82 & 0.84 & 0.35 & 0.36 & 0.35 \\
\hline
\end{tabular}
}
\end{center}
\caption{Performance of seq2slate and other baselines on data generated with similar-clicks.}
\label{tab:letor_similarity}
\end{table*}

\begin{table*}[t]
\begin{center}
{\footnotesize
\resizebox{0.95\textwidth}{!}{%
\begin{tabular}{|c|cccc|cccc|}
\hline
\multirow{2}{*}{Ranker} & \multicolumn{4}{|c|}{Yahoo} & \multicolumn{4}{|c|}{Web30k} \\
{} & MAP & NDCG@5 & NDCG@10 & rank-gain & MAP & NDCG@5 & NDCG10 & rank-gain \\
\hline
seq2slate & \bf{0.82} & \bf{0.82} & \bf{0.84} & \bf{8.5} & \bf{0.44} & \bf{0.54} & \bf{0.50} & \bf{16.0} \\
Greedy policy      & \bf{0.82} & \bf{0.82} & \bf{0.84} & \bf{8.5} & 0.44 & 0.54 & 0.50 & 15.9 \\
smooth-hinge        & 0.80 & 0.80 & 0.82 & 7.7 & 0.44 & 0.54 & 0.50 & 15.9 \\
\textsc{reinforce}  & \bf{0.82} & \bf{0.82} & \bf{0.84} & \bf{8.5} & 0.42 & 0.53 & 0.49 & -14.8 \\
one-step decoder    & 0.81 & 0.81 & 0.82 & 7.7 & 0.44 & 0.53 & 0.49 & 15.5 \\
shuffled data       & 0.79 & 0.78 & 0.79 & -- & 0.42 & 0.48 & 0.46 & -- \\
base ranker (no-op) & 0.78 & 0.76 & 0.79 & 0 & 0.43 & 0.53 & 0.49 & 0 \\
\hline
\end{tabular}
} 
} 
\end{center}
\caption{Comparison of model and data variants for seq2slate on data generated with similar-clicks.}
\label{tab:letor_similarity_variants}
\end{table*}

\paragraph{Adaptivity to the type of interaction}
To demonstrate the flexibility of seq2slate, we generate data using
a variant of the diverse-clicks model above.
Specifically, in the \emph{similar-clicks} model, the user also clicks on observed irrelevant items if they are similar to previously clicked items (increasing the number of total clicks).
As above,
we use the pairwise distances in feature space $D_{ij}$ to determine similarity.
For this model we use $q=0.5$, and $\eta=0.3$ for Web30k, $\eta=0.1$ for Yahoo, to keep the proportion of positive labels similar.%
\footnote{The value of $\eta$ was chosen such that the percentage of examples with no positive labels (clicks) at all remained small enough and roughly the same in all datasets (around 1.15\% of all examples).}
The results in \tabref{tab:letor_similarity} show that seq2slate has comparable performance to the baseline rankers, with slightly lower performance on Yahoo and significantly better performance on the harder Web30k data.
This demonstrates that our model can adapt to various types of interactions in the data.
Notice that no changes to the model or training algorithm were necessary for seq2slate. In contrast, if one used a specific interaction model for `diverse-clicks', then a different model would be required for the `similar-clicks' data, a distinction not needed with seq2slate.

\subsection{Real-World Data}
\label{sec:real_experiments}
We also apply seq2slate to a ranking problem from a large-scale commercial recommendation system.
We train the model using massive click-through logs (comprising roughly $O(10^7)$ instances) with cross-entropy loss, the greedy policy,
L2-regularization
and dropout.
The data has item sets of varying size, with an average $n$ of 10.24 items per example.
We learn embeddings of the raw inputs as part of training.

\tabref{tab:real_data} shows the performance of seq2slate and the one-step decoder compared to the production base ranker on test data
(of roughly the same size as the training data).
Significant gains are observed in all performance metrics, with sequential decoding outperforming the one-step decoder. This suggests that sequential decoding may more faithfully capture complex dependencies between the items.

\begin{table}[t!]
\centering
{\small
\begin{tabular}{|c|c|c|c|c|}
\hline
Ranker & MAP & NDCG@5 & NDCG@10 & rank-gain \\
\hline
one-step decoder & +26.79\% & +10.69\% & +40.67\% & 0.83 \\
seq2slate & +31.32\% & +14.47\% & +45.77\% & 1.087 \\
\hline
\end{tabular}
}
\caption{Performance compared to a competitive base production ranker on real data.}
\label{tab:real_data}
\end{table}

\begin{figure}[t]
  \centering
  \includegraphics[width=200pt]{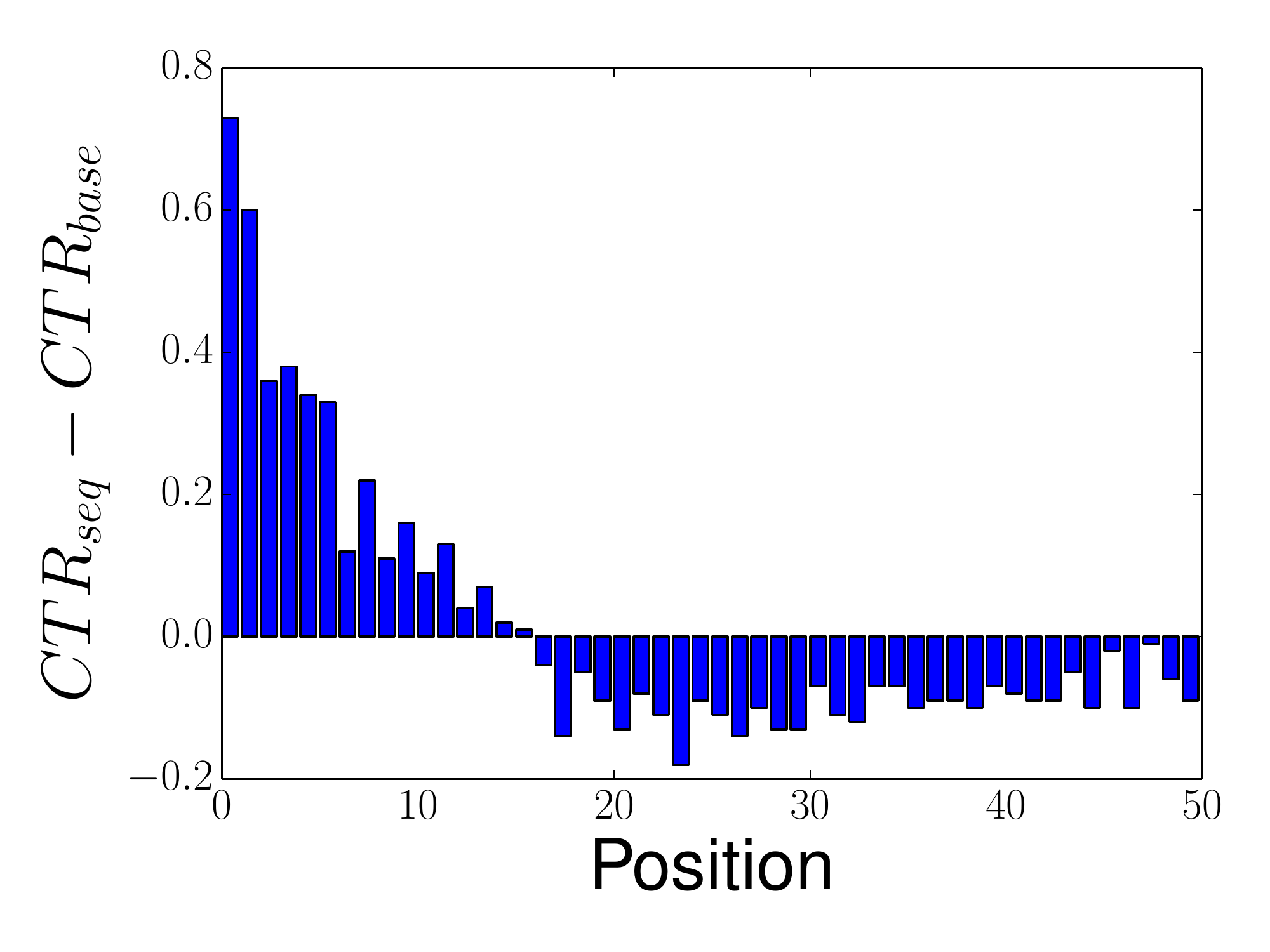}
  \caption{Difference in CTR per position between a seq2slate model and a base production ranker in a live experiment.}
  \label{fig:rank_dist}
\end{figure}

Finally, we let the learned seq2slate model run in a live experiment (A/B testing) and re-rank the result of the current production recommender system.
We compute the click-through rate (CTR) in each position (\#clicks/\#examples) for seq2slate.
The production base ranker serves traffic outside the experiment, and we compute CTR per position for this traffic as well.
\figref{fig:rank_dist} shows the difference in CTR per position, indicating that seq2slate has significantly higher CTR in the top positions. This suggests that seq2slate indeed places items that are likely to be chosen higher in the ranking.

\section{Related Work}
In this section we discuss additional related work. We build on the recent impressive success of seq2seq models in complex prediction tasks, including
machine translation \cite{Sutskever2014,Bahdanau2015},
parsing \cite{vinyals2015parsing},
combinatorial optimization \cite{vinyals2015,bello2017nco},
multi-label classification \cite{wang2016cnn-rnn,nam2017multilabel},
and others.
Our work differs in that we explicitly target the ranking task, which requires a novel approach to training seq2seq models from weak feedback (click-through data).

Most of the work on ranking mentioned above uses shallow representations. However, in recent years deep models have been used for information retrieval, focusing on embedding queries, documents and query-document pairs \cite{huang2013,Guo2016cikm,palangi2016,Wang2017attnletor,Pang2017DeepRank} (see also recent survey by \citet{mitra2017}).
Rather than embedding individual items, in seq2slate a representation of the entire slate of items is learned and encoded in the RNN state.
Moreover, learning the embeddings ($x$) can be easily incorporated into the training of the sequence model to optimize both simultaneously end-to-end.

Closest to ours are the recent works of \citet{Mottini2017kdd} and \citet{ai2018sigir}, where an RNN is used to encode a set of items for ranking.
There are some differences between the approach of \citet{ai2018sigir} and ours, including using GRU cells instead of LSTM cells, reversing the input order (the highest ranking item is fed to the encoder last), and training from relevance scores instead of click-through data.
More importantly, both works \cite{Mottini2017kdd,ai2018sigir} use a single decoding step.
In contrast, we apply sequential decoding, which directly allows item scores to change based on previously chosen items.
We believe that this significantly simplifies modeling and inference with complex high-order interactions between items, and indeed show that it performs much better
in practice (see \secref{sec:real_experiments}).

Finally, \citet{SantaCruz:CVPR2017} recently proposed an elegant deep learning framework for learning permutations based on the so called Sinkhorn operator, building on prior work by \citet{adams2011sinkhorn}. Their approach uses a continuous relaxation of permutation matrices (i.e., the set of doubly-stochastic matrices, or the Birkhoff polytope).
Followup work has focused on improved training and inference procedures, including a Gumbel softmax distribution to enable efficient learning \cite{mena2018},
a reparameterization of the Birkhoff Polytope for variational inference \cite{linderman18a},
and an Actor-Critic policy gradient training procedure \cite{emami2018}.
However, these works are focused on reconstruction of scrambled objects (i.e., matchings), and it is not obvious how to extend it to our ranking setting, where no ground-truth permutation is available.

\section{Conclusion}
We presented a novel approach to ranking sets of items called seq2slate.
We found the formalism of pointer-networks particularly suitable for this setting.
We emphasized the modeling and computational advantages of using sequential decoding, which allowed the model to dynamically adjust placement of items on the slate given previous choices.
We addressed the challenge of training the model from weak user feedback (click-trough logs) to improve the ranking quality.
To this end, we proposed new sequence losses along with corresponding gradient-based updates.
Our experiments show that the proposed approach is highly scalable and can deliver significant improvements in ranking results.

Our work can be extended in several directions.
In terms of architecture, we aim to explore the \emph{Transformer} network \cite{vaswani2017transformer,universal_transformers} in place of the RNN.
Several algorithmic variants can potentially improve the performance of our model.
For inference, beam-search has been shown to improve predictions of several seq2seq models \cite{wiseman2016}, and we believe can do the same for seq2slate.
For training, several approaches have been recently proposed for seq2seq models, including Actor-Critic \cite{Bahdanau2017actor_critic} and more recently SeaRNN \cite{searnn2018leblond}, and it will be interesting to test their performance in the ranking setting.

Finally, an interesting future direction is to study off-policy correction for seq2slate \cite{joachims2018,minminWSDM19}.
In this setting, training examples are assigned \emph{importance weights} in order to account for the fact that the labels were obtained using a different policy than the one we wish to evaluate during training.
In particular, the expected sequence loss is adjusted to account for this mismatch as follows:%
\footnote{Substituting $\mathcal{L}_\pi(\theta)$ by $\mathcal{R}(\pi,y)$ yields an equivalent formulation for the expected reward from \secref{sec:policy_gradient}.}
\[
\mathbb{E}_{\pi \sim p_{\theta}(.|x)} [{\mathcal{L}}_\pi(\theta)]
=
\mathbb{E}_{\pi \sim p_{\text{base}}(.|x)}
\left[\frac{p_{\theta}(\pi|x)}{p_{\text{base}}(\pi|x)}
      {\mathcal{L}}_\pi(\theta)\right]~,
\]
where $p_\text{base}$ is the probability of $\pi$ under the base ranker (i.e., logging policy).
This expectation can then be approximated from logged samples as in \secref{sec:training}.
We leave this extension to future work.

\appendix

\section{Derivation of the Expected Loss}
\label{app:expected_loss}
Here we show the expected loss as a function of the model scores $S$,
{\small
\begin{align*}
\expect[{\mathcal{L}}(\theta)] =&~ \sum_\pi p(\pi) {\mathcal{L}}_\pi(\theta) \\
=&~
\sum_\pi p(\pi) \sum_j \ell_{\pi_{<j}}(\theta) \\
=&~
\sum_j \sum_\pi p(\pi_{<j}) p(\pi_{\ge j}|\pi_{<j}) \ell_{\pi<j}(\theta) \\
= &~
\sum_j \sum_{\pi_{<j}} p(\pi_{<j}) \ell_{\pi_{<j}}(\theta) \overbrace{\sum_{\pi_{\ge j}} p(\pi_{\ge j}|\pi_{<j})}^{1} \\
= &~
\sum_j \sum_{\pi_{<j}} \left( \prod_{k=1}^{j-1} e^{s^k_{\pi_k}} / \sum_{i\notin \pi_{<k}} e^{s^k_i} \right) \ell_{\pi<j}(s^j,y) ~~.
\end{align*}
}

\clearpage
\bibliographystyle{plainnat}

\end{document}